\documentclass[a4paper,10pt]{article}
\usepackage{epsfig,amsmath,amsfonts}%
\newcommand{\eqsection}{\makeatletter
    \@addtoreset{equation}{section}
    \renewcommand{\theequation}{\arabic{section}.\arabic{equation}}
    \makeatother}

\def\lal{&&\nqq {}}

\def\beq{\begin{equation}}
\def\eeq{\end{equation}}
\def\bear{\begin{eqnarray}}
\def\bearr{\begin{eqnarray} \lal}
\def\ear{\end{eqnarray}}
\def\earn{\nonumber \end{eqnarray}}

\def\dst{\displaystyle}
\def\tst{\textstyle}
\def\fracd#1#2{{\dst\frac{#1}{#2}}}
\def\fract#1#2{{\tst\frac{#1}{#2}}}
\def\Half{{\fracd{1}{2}}}
\def\half{{\fract{1}{2}}}

\def\const{{\rm const}}

\begin{document}

\begin{center}
\LARGE{\bf {Chiral Cosmological Models: Dark Sector Fields Description}} 
\end{center}

\begin{center}

Chervon S.V. \\%

\sl{\it {Laboratory of Gravitation, Cosmology, Astrophysics,\\
 Ilya Ulyanov State Pedagogical University, Ulyanovsk 432700, Russia}}

\end{center}
\small{{\underline{ Abstract.}} The present review is devoted to a
Chiral Cosmological Model as the self-gravitating nonlinear sigma model with
the potential of (self)interactions employed in cosmology. The chiral
cosmological model has
successive applications in descriptions of the inflationary epoch
of the Universe evolution; the present accelerated expansion of
the Universe also can be described by the chiral fields multiplet
as the dark energy in wide sense.

To be more illustrative we are often addressed to the
two-component chiral cosmological model. Namely, the two-component
chiral cosmological model describing the
phantom field with interaction to a canonical scalar field is
analyzed in details. New generalized model of quintom character is
proposed and exact solutions are founded out.

In the review we
represented the perturbation theory for chiral cosmological model
 with the aim to
describe the structure formation using the progress achieved in
the inflation theory.
It was shown that cosmological perturbations from chiral fields can be
decomposed for inflaton and
the dark sector perturbations. The two-component model is investigated
in details, the general solution for shortwave approximation is
obtained and analyzed for power law Universe expansion.

New issue for understanding the features of Universe evolution is
proposed by consideration of the dark sector fields on the
inflaton background. The results are illustrated for the solutions
in the long-wave approximation.
}

\section{Introduction}

Active investigations of the Early Universe have been originated with understanding
of inflationary epoch necessity in the beginning of the 80-s last century. During the decade
it became clear strong relation between particle physics (micro-physics) and cosmology
(macro-physics) leading to connection
of quantum fluctuations in the very early Universe with it large scale structure observed nowadays.
Such topics as topological defects, inflation, dark matter, axions and quantum cosmology were studied
by many scientists. The questions about physical roots of the Universe expansion source
(inflation) or quantum creation of the Universe were discussed in the literature as well
(for review see \cite{koltur90}).

The discovery of the current acceleration of the Universe at the end of the 20-th century
leads to a great number of
new theoretical hypothesis to explain this observational fact. We can think over Dark
Energy (DE) as the source of modern accelerated expansion of the Universe in a wide sense.
The most often DE is represented by the cosmological constant $\Lambda $, quintessence
as a canonical scalar field with the potential or as the scalar fields coupled to other
species of matter/radiation and many others (see, for example, \cite{cosats06}).
The general term
{\it dark sector fields} is used to cover all kinds of DE and Dark Matter (DM).
let us
study a chiral nonlinear sigma model coupled to gravity in order to make the general
framework for dark sector
fields and take into account some generalization for multiple-field model.

A self-gravitating Nonlinear Sigma Model (NSM) with a potential
of (self)interactions has been applied for description of inflationary
cosmology
\cite{Chervon95b} and therefore was called {\it The Chiral Inflationary Model}.
Two years earlier pure kinetic NSM has been considered in cosmology \cite{Chervon93},
but it could not carry all necessary properties of inflationary
period because the equation of state was corresponded to ultra-stiff  matter only.

The term "chiral" is considered in the sense of short equivalent to the general chiral
NSM with a Riemannian metric as a target space in accordance with terminology
in the review \cite{Perelomov87}. Paradoxically, this term has been applied to pure
bosonic NSM which has no relation to chiral symmetry because pure bosonic NSM
does not contain the spinor fields. Thus the term "chiral" means that
the scalar fields are not free and take values on some nonlinear manifold.

\subsection{NSM in Cosmology}

The chiral cosmological model (CCM) is based on the NSM with the potential of (self)interactions
and CCM are generalization for a theory of scalar field singlet which has a lot of applications
in cosmology. Therefore it is of interest to consider cosmological applications of the NSM.

NSM coupled to the gravitational field has been introduced from the different positions.
The 4-dimensional NSM as a generalization from 2-dimensional one can be presented
only with coupling to the gravitational field if one wants to have desirable analogy
with the gauge theories expressed in existence of instanton and meron solutions. This
approach (based on a Riemannian space-time with Euclidian signature) has been realized
by De 'Alfaro et al \cite{dealfaro79}. %

Considering the NSM as the source of the gravitational field (based on a Riemannian
space-time with Lorentzian signature) George Ivanov \cite{Ivanov83}
independently of work \cite{dealfaro79} proposed the self-gravitating
NSM as a generalization from multiple scalar fields theory. In the work \cite{Ivanov83}
the basic equations of the self-gravitating NSM were derived and the methods of solving them
were suggested. The applications of the suggested methods for spherical-symmetric,
plane-symmetric space-times  and for cosmological metrics have been demonstrated and
examples of exact solutions were obtained. The cosmological solutions include:
A) SO(4)-invariant solution for closed Friedmann -- Robertson -- Walker Universe with a
constant scale factor $a(t)= a_0$ = const (a static universe) or with $a(t) \propto t $ with
effective Equation of State (EoS) corresponding to the radiation $p=-\frac{\rho}{3}$;
B) SO(3)-invariant solution for an anisotropic universe.

Let us note here that mentioned above cosmological solutions have been obtained
with the isometric embedding method also proposed for the self-gravitating NSM in the
work \cite{Ivanov83}.%

Further applications of NSM in cosmology were started from the works \cite{Chervon92},\cite{Chervon93}.
In parallel with our investigations (Chervon et al, 1992-2012) there were few works
on the NSM applications in cosmology. In the work \cite{Jaffe94} the evolution of
perturbations to the flat FRW universe has been considered for $O(N)$ nonlinear
sigma model at large $N$. The feature of the proposed model was the inclusion into consideration
the "stiff source" which does not affect on the background determined by multicomponent
perfect fluid. The effect of the "stiff source" is compared with metric and density of fluid
perturbations; the stiff source may lead to initially non-Gaussian distribution
on scales of several hundreds megaparsecs. Further extension of this model for the inflationary
epoch and analysis of influence of the stiff source on the large scale structure formation
was presented at the work \cite{chepan10}.

Multiple scalar fields coupled to gravity have been considered in the framework of
scalar-tensor theories of gravitation in the work \cite{berhel94}. 
It was stated out
that for realization of the exact solar system tests it was necessary that some scalar kinetic terms should be non-positive-definite. This statement means the existence of phantom fields
in the Universe.

The spherically-symmetric, static solutions to the $SU(2)$ NSM on de Sitter background
have been studied by numerical methods in the work \cite{aiclec97}.
It was found that the countable set of regular solutions with finite
energy exists there.

\subsection{Coupled dark sector fields}

After the discovery of the Universe acceleration the concept of Dark Energy (DE)
has been actively studied.
First of all DE is considered as the cosmological constant and the scalar field
(quintessence, phantom, tachyon etc.). Other presentations of DE include modified
gravity, Chaplygin gases and many other models \cite{cosats06}. 

The goal of our work is to generalize the models with interacting species of the
Universe of the scalar field type and to describe them in terms
of the chiral cosmological model.
We will not pay much attention here to interaction of the scalar field with cosmic fluid or
dust/radiation/ordinary matter.

The authors of the work \cite{sahwan99}  
proposed the new cosmological model as
a unified picture of quintessence and new form of DM. This, so called Frustrated CDM
model, deals with two gravitationally interacting scalar fields without kinetic and
potential interaction. Further application of this model to reconstruction quintessence
from supernova observations is  considered in \cite{Sahni01}.

It was proposed in the works \cite{matos00a}, \cite{matos00b} 
the cosmological model
in which  DM and DE are modelled by two gravitationally coupled scalar fields . This model
contains 95\% of scalar fields. The shape of the potentials was chosen in specific form
which gave opportunity to define the scalar DM mass (of order $10^{-26}$ eV). The
implication on the structure formation was analyzed and the minimal cat-off radio
($r_c \sim 1.2 $ kpc) for this model was obtained.

The models with a nonminimal coupling between
DM and scalar field DE were investigated in the article \cite{Koivisto05}.
The evolution of linear perturbations
for such models was considered in details. It should be mention that proposed
nonminimal coupling may lead for kinetic interaction if DM specie will be presented as
the scalar field.

It was first time mention \cite{canood06} 
about the fact that the
same scalar field can play the role of early-time
inflaton and later-time DE. The (dark) matter can be included into consideration by
an ordinary way as a cosmic fluid.

The article \cite{zhao05}  
is devoted to observation effects from quintom models.
It should be noted that in special cases quintom models can be presented with two
gravitationally and potentially interacting scalar fields.

Investigation of the dynamic interaction
between DE and DM have been analyzed in \cite{bohmer08}. 
The general coupling between  DE and DM is described by the additional term
in the energy balance equation. This type of interaction can be extended for two
gravitationally coupled scalar fields if DM will be presented as a scalar field. Note
that energy transfer DM to DE and DE to DM was considered in the work
\cite{valiviita08} 
as well.

The model with N-dimensional internal space
for a multi-field configuration was proposed in the work \cite{chimento09}.
The difference from the chiral cosmological model \cite{Chervon95b}, \cite{Chervon95a}
is in the restriction
of the internal space metric for the constant one.

The work \cite{setsar08} 
is devoted to quintom model with $O(N)$ symmetry.
The model is of importance
because it presents the nonlinear sigma model with deformed chiral space
\cite{Chervon95b}, \cite{Chervon95a}. Moreover the article \cite{setsar08}
may be considered as
the next step from the work \cite{chimento09} 
on the way to the chiral cosmological
models (CCM): it presents CCM with the special choice of the internal space metric.

The dynamical properties of DE model with two gravitationally, kinetically
and potentially coupled scalar fields were studied in the work \cite{vandebruck09}.
The authors started from consideration of two-fields model with  noncanonical
kinetic term. This model is more wide then CCM and can be reduced for it under simple
choice of the action. In the mentioned article the focus was exactly on the described
situation: the chosen model was equivalent to the 2-component CCM with the cross
interaction between scalar (chiral) fields. Further extension of the work
\cite{vandebruck09} is presented in \cite{sarwel09}. %

\section{Chiral cosmological model}\label{CCM}

The action of CCM as the the
action of the self-gravitating nonlinear sigma model (NSM) with the
potential of (self)interactions $V(\varphi)$ \cite{Chervon95b}, \cite{Chervon95a}
 and cosmological constant $ \Lambda$ reads:
\begin{equation}\label{1}
  S=\int\sqrt{-g}d^4x\left(\frac{R-2\Lambda}{2\kappa}+
    \frac{1}{2}h_{AB}(\varphi)\varphi^A_{,\mu}\varphi^B_{,\nu}g^{\mu\nu}-V(\varphi)\right),
\end{equation}
where $ R $ being a scalar curvature of a Riemannian manifold with the metric
$g_{\mu\nu}(x)$, $\kappa$ -- Einstein gravitational constant, $\varphi=(\varphi^1,\ldots,\varphi^N)$~being a
multiplett of the chiral fields (we use a notation
$\varphi^A_{,\mu}=\partial_{\mu}\varphi^A =\frac{\partial
  \varphi^A}{\partial x^\mu}$),
$h_{AB}$~being the metric of the target space (the chiral space) with
the line element
\beq\label{tsmet}
 d s^2_{\sigma}=h_{AB}(\varphi)d\varphi^A d\varphi^B,~~A,B,\ldots =\overline{1,N}.
\eeq
The energy-momentum tensor for the model (\ref{1}) reads
\begin{equation}\label{ch-em}
  T_{\mu\nu}=\varphi_{A,\mu}\varphi^A_{,\nu}-
  g_{\mu\nu}\left(\frac{1}{2}\varphi^A_{,\alpha}\varphi^B_{,\beta}g^{\alpha\beta}h_{AB}-
    V(\varphi)\right).
\end{equation}
The Einstein equation can be transformed to
\beq\label{ein}
R_{\mu\nu}=\kappa\{
h_{AB}\varphi^A_{,\mu}\varphi^B_{,\nu}- g_{\mu \nu} V(\varphi)\}
- \Lambda g_{\mu\nu}
\eeq
which simplify the derivation of gravitational dynamics equations.

The standard form of the Einstein equation
\beq\label{ein-st}
R_{\mu\nu} -\half g_{\mu\nu} R -\Lambda
g_{\mu\nu} = \kappa T_{\mu\nu},~~\kappa=8\pi G
\eeq
will be also under
consideration. 

Varying the action (\ref{1}) with respect to $\varphi^C$, one can
derive the dynamic equations of the chiral fields
\begin{equation}\label{4}
  \frac{1}{\sqrt{-g}}\partial_{\mu}(\sqrt{-g}\varphi^{,\mu}_A)-\frac{1}{2}\frac{\partial
    h_{BC}}{\partial\varphi^A}\varphi^C_{,\mu}\varphi^{B}_{,\nu}g^{\mu\nu}+V_{,A}=0,
\end{equation}
where $V_{,A}=\frac{\partial V}{\partial\varphi^A}$.
Considering the action (\ref{1}) in the framework of
cosmological spaces, we arrive to a chiral (inflationary or) cosmological model
\cite{Chervon95b}, \cite{Chervon95a}, \cite{ch00mg}, \cite{ch02gc}.

\subsection{The two-component chiral cosmological model}

To connect our consideration with dark sector fields interaction models
presented in the Introduction let us start from analysis of the two-component
CCM.
The target space metric we select in the most general form taking into account
nondiagonal metric component $h_{12}$

\beq\label{tsmet}
ds^2_{\sigma}=h_{11}(\phi,\psi)d\phi^2 +2h_{12}(\phi,\psi)d\phi
d\psi+h_{22}(\phi,\psi)d\psi^2.
\eeq

Thus we have two component chiral cosmological
model as the source of the gravitational field with the target space
metric (\ref{tsmet}) and two chiral fields denoted as
$$
\varphi^1=\phi,~~\varphi^2=\psi.
$$

In terms of chosen target space (\ref{tsmet}) the energy-momentum
tensor (\ref{ch-em}) can be presented in the following form
\beq\label{em-2}
T_{\mu\nu}=h_{11}\phi_{,\mu}\phi_{,\nu}+2h_{12}\phi_{,\mu}\psi_{,\nu}+
h_{22}\psi_{,\mu}\psi_{,\nu}-
g_{\mu\nu}\left[ \Half h_{11}\phi_{,\rho}\phi^{,\rho}+
h_{12}\phi_{,\rho}\psi^{,\rho}+ \Half h_{22}\psi_{,\rho}\psi^{,\rho}-
V(\phi,\psi)\right].
\eeq

The metric of homogeneous and isotropic Universe we take in the
Friedman -- Robertson -- Walker (FRW) form
\begin{equation}\label{frw}
  d s^2=d t^2-a(t)^2\left(\frac{d r^2}{1-\epsilon r^2}+r^2d\theta^2+r^2\sin^2\theta
    d\varphi^2\right).
\end{equation}

The Einstein equation with $\Lambda =0 $ can be represented in the form
\bear\label{E1}
H^2=\frac{\kappa}{3}\left[\Half h_{11}\dot\phi^2 +h_{12}\dot\phi \dot\psi+
\Half h_{22}\dot\psi^2 +V(\phi,\psi) \right]-\frac{\epsilon}{a^2}, \\
\label{E2} \dot H =-\kappa \left[ \Half h_{11}\dot\phi^2
+h_{12}\dot\phi \dot\psi+\Half h_{22}\dot\psi^2\right]+\frac{\epsilon}{a^2},
\ear
where the overdot means the derivative with respect to cosmic time $t$.

The chiral fields equations for the model (\ref{tsmet})
in FRW universe (\ref{frw}) reads
\bear\label{F1}
3H\left( h_{11}\dot\phi +h_{12}\dot\psi \right)  + \partial_t\left( h_{11}\dot\phi
+h_{12}\dot\psi \right) -\half \frac{\partial h_{11}}{\partial \phi}\dot\phi^2 -
\frac{\partial h_{12}}{\partial \phi}\dot\phi \dot\psi-
\half \frac{\partial h_{22}}{\partial \phi}\dot\psi^2 +\frac{\partial V}{\partial\phi}=0, \\
\label{F2}
3H\left( h_{12}\dot\phi +h_{22}\dot\psi \right)  +\partial_t\left( h_{12}\dot\phi
+h_{22}\dot\psi \right)  -\half \frac{\partial h_{11}}{\partial \psi}\dot\phi^2 -
\frac{\partial h_{12}}{\partial \psi}\dot\phi \dot\psi-
\half \frac{\partial h_{22}}{\partial \psi}\dot\psi^2 +\frac{\partial V}{\partial\psi}=0.
\ear

Let us mention that one of the Einstein equations -- Friedmann equation (\ref{E1}) can
be obtained by integrating the linear combination of the fields equations
$\dot\phi $ (\ref{F1}) +$\dot\psi $ (\ref{F2}) using the
second Einstein equation (\ref{E2}).

If the law of Universe evolution   $a=a(t)$ is specified the system of equations (\ref{E1})-(\ref{F2}) can be considered as the system
of differential equations of the second order with three unknown functions: two chiral
fields $\varphi$ and $\psi$, and the potential $V$.
According with the method of fine tuning of the potential \cite{chzhsh97} 
we consider the scalar factor as given function on cosmic time : $a=a(t)$.
The metric of a target space we will not fix as it traditionally accepted in HEP, leaving
a freedom of adaptation to resolving problem. Making simple
algebraic conversion of Einstein equations (\ref{E1})--(\ref{E2})
one can find their useful implication:
\begin{equation}\label{12}
  \frac{1}{2}h_{11}\dot\phi^2(t)+h_{12}\dot\phi (t)\dot\psi (t) +\frac{1}{2}h_{22}(t)\dot\psi^2(t)=
  \frac{1}{\kappa}\left[\frac{\epsilon}{a^2}-\dot
    H\right],
\end{equation}
\begin{equation}\label{13}
  V(t)=\frac{3}{\kappa}\left(H^2+\frac{1}{3}\dot
    H+\frac{2}{3}\frac{\epsilon}{a^2}\right).
\end{equation}

For the sake of completeness let us present the consequence of the Einstein equation in the
form often used for analysis of Universe acceleration parameter 

\beq
\frac{\ddot{a}}{a}=-\frac{\kappa}{3}\left[ \frac{1}{2}h_{11}\dot\varphi^2(t)+
h_{12}\dot\varphi (t)\dot\psi (t) +\frac{1}{2}h_{22}(t)\dot\psi^2(t) -V\right].
\eeq

Let us turn our attention to the models with two scalar fields and possible
kinetic interactions between them. This models are often called as
{\it multicomponent or multiple-field inflationary}
models which clearly \cite{ch97b} are contained into {\it chiral cosmological models}.
Multiple-fields models describing dark sector fields we can divide into the following
classes.

{\it I Class} -- the models with two canonical scalar fields
with different physical meaning of the fields. {\it II Class} -- quintom models
with one phantom and one canonical scalar fields. {\it III Class} -- are the models with
kinetic interactions between scalar fields. It is clear that the {\it I Class} can
be described with $h_{11}=1,~h_{22}=1,~h_{12}=0$;  the {\it II Class} can
be described with $h_{11}=1,~h_{22}=-1,~h_{12}=0$  and the {\it III Class} can
be described with $h_{11}=1,~h_{22}= 1,~h_{12}=f(\phi, \psi).$ Let us note that
the case with $h_{11}=1,~h_{22}= -1,~h_{12}=f(\phi, \psi)$ has not been considered
in the literature yet. 

On the other hand we have possibility to analyze the target space metric $h_{AB} $
as the two
dimensional surface $\Sigma $ embedding into appropriate three dimensional space.
Then in according
with the surface's points classification we arrive to the following three classes.
Namely, if the determinant of the chiral space metric
$h = det (h_{AB})=h_{11}h_{22} - h_{12}^2 $ has positive sign: $h>0 $ at the point
$M \in \Sigma $, we have the {\it elliptic } point; if $h<0$ -- the {\it hyperbolic}
point and if $h=0$ with at least one from metric coefficient $h_{AB} \neq 0 $
we have the {\it parabolic} point.

Thus our simple division above may be characterized by elliptic points (I-st class),
hyperbolic points (II-nd class). The III-d class requests an additional investigation.
If we consider the case with $h=0$ and {\it all metric components does not equal
to zero} we can reduce the model to one canonical field as it was done in the work
\cite{vandebruck09}  
for the case $h_{11}=1,~h_{22}= 1,~h_{12}=\frac{a}{2}$
(with $a=constant $) when $a=\pm 2$. Namely,  we can introduce new field $\Theta $ by
the relation
$$
\Theta = \sqrt{h_{11}}\phi \pm \sqrt{h_{22}}\psi + \Theta_*,~~ (\Theta_*=const).
$$
Then the cosmological dynamics will be reduced to the self-gravitating scalar singlet
$\Theta $.

Summing up the arguments above we can conclude that the chiral cosmological model
gave for us good possibility not only describe the classes investigated in the
literature mentioned above but
also to consider the evolution of kinetic interactions between scalar fields. To this
end we may introduce the dependance of the chiral metric
components on the scalar fields:
$h_{11}=h_{11}(\phi,\psi),~h_{12}=h_{12}(\phi,\psi),~h_{22}=h_{22}(\phi,\psi).$
Thus we make generalization of the multi-fields models
and now, using the freedom in the choice of the chiral metric, we can search for
evolutionary solutions and compare them with observation data. Let us also note that
the investigation of possible functional dependence of the chiral metric components
may lead to
better understanding of physically important scalar fields governing the Universe
evolution.

\subsection{The generalized quintom models}\label{gqm}

Let us consider new parametrization of the quintom model. Namely, let us set
$$
h_{11}=-1,~~h_{22}=h_{22}(\phi),~~h_{12}=0,
$$
where $h_{22}$ is a given function on the first chiral field $\phi$.
This approach means that we have fixed the metric of internal (chiral) space,
or by other words -- the symmetry of the chiral space. The original chiral nonlinear
sigma model belongs to this category as having $SU_2 \times SU_2$ symmetry.

Let us consider quintom analog to $SO(3)$ nonlinear sigma model, which has (in two
dimensions)
an interesting properties making the analogy with gauge theories: instanton and
meron solutions, asymptotical freedom etc.
To derive the system of equations (\ref{E1})-(\ref{F2}) for quintom generalization
of $SO(3)$ NSM we choose the following chiral metric components:
\beq
h_{11}=-1,~~h_{22}=\sin^2\phi,~~h_{12}=0.
\eeq

After the substitution one can obtain the following system of equations
\bear\label{E1-q}
H^2=\frac{\kappa}{3}\left[-\Half \dot\phi^2 +
\Half \sin^2\phi \,\dot\psi^2 +V(\phi,\psi) \right]-\frac{\epsilon}{a^2}, \\
\label{E2-q} \dot H =-\kappa \left[-\Half \dot\phi^2
+\Half \sin^2\phi\,\dot\psi^2\right]+\frac{\epsilon}{a^2},\\
\label{F1-q}
-\ddot{\phi}-3H\dot\phi -
\half \sin 2\phi\,\dot\psi^2 +\frac{\partial V}{\partial\phi}=0, \\
\label{F2-q}
\ddot{\psi}+3H\dot\phi +
2\cot \phi\,\dot\phi\, \dot\psi + \sin^{-2}\phi\, \frac{\partial V}{\partial\psi}=0.
\ear

To simplify the system above let us consider a spatially-flat Universe with
$\epsilon =0$. Also we may suggest that the potential of (self)interaction can
be considered as the constant during some period of Universe evolution in analogy with
inflationary period, when the potential is considered as plane. From the other hand the
same situation we can consider as kinetic model but with cosmological term.
Thus we denote $ V=\const =\Lambda $. Also let us put the gravitational constant $\kappa$
equal to unity: $\kappa =1$. After that one can reduce from combination of the equations
(\ref{E1-q}) and (\ref{E2-q}) the equation on Hubble parameter $H$:
\beq\label{H-1}
3H^2+\dot{H} = \Lambda .
\eeq
The well-known solutions are
\beq\label{H-th}
H =\sqrt{\frac{\Lambda}{3}}\tanh (\sqrt{3\Lambda}t),
\eeq
\beq\label{a-ch}
a=a_*[\cosh (\sqrt{3\Lambda}t)]^{1/3}.
\eeq
Besides, if we request the density $\rho =K+V = K + \Lambda $ should be positive
then $\Lambda > 0$ and $\dot{H}<\Lambda $. The last expression is always hold for the
solution (\ref{a-ch}).

Let us consider the evolution of equation of state
$$
\omega_{DS}=\frac{K-\Lambda}{K+\Lambda}.
$$

Using the equations (\ref{12})-(\ref{13}) and the solution (\ref{a-ch}) we can
obtain the following asymptotes
$$
\omega_{t \rightarrow 0} \rightarrow -\infty, ~~\omega_{t \rightarrow \infty} \rightarrow -1.
$$

Thus we can state the strong phantom influence at the early stage of the Universe
evolution and domination of DE at the late stages.

Integrating the equation (\ref{F2-q}) one can obtain the relation
\beq\label{q-psi}
\dot{\psi}=\frac{C_1}{\cosh (\sqrt{3\Lambda}t)\sin^2\phi},~~C_1=\const.
\eeq
With this relation we can integrate equation (\ref{F1-q}). 
The solution for the first (phantom) field can be obtained from the expression
\beq\label{q-phi-I}
\cos \phi = -\frac{\sqrt{C_1^2+2 \Lambda}}{\sqrt{2\Lambda}}
\sin \left(\sqrt{\frac{2}{3}}\arctan\left(\sinh (\sqrt{3\Lambda}t)\right)+C_2\right).
\eeq

It is worth to note that the solution (\ref{q-phi-I}) gives some restriction for
the constant $C_1$.

The second (canonical) field can be obtained from (\ref{q-psi}).
The solution is
\bear
\psi -\psi_0 = \frac{1}{2a_*}\left[\ln \left|\frac{C_1}{\sqrt{2\Lambda}}\tan z+1\right|
-\ln \left|\frac{C_1}{\sqrt{2\Lambda}}\tan z-1\right|\right],\\
z=\sqrt{\frac{2}{3}}\arctan (\sinh (\sqrt{3\Lambda}t))+C_2.
\ear
With the same approach we can find the solutions for $ h_{22}=\sinh^2 \phi $ and
other interesting cases including the changing of the fantom and canonical
fields in the chiral metric \cite{cheabb12}.
For example the case with
$h_{22} = \phi^2 $ gives the analytic solution
\bear
\phi = \pm \sqrt{2z^2-\frac{C_1^2}{2a_*^3\Lambda}},~~z=
\arctan\left(\sinh(\sqrt{3\Lambda}t)\right)+C_2, \\
\psi = \frac{a_*^3}{2}\left[\ln \left(\frac{C_1^2\sqrt{3}}{2a_*^3
\sqrt{\Lambda}}-z\right)-\ln \left(\frac{C_1^2\sqrt{3}}{2a_*^3
\sqrt{\Lambda}}+z\right)\right].
\ear

Thus we can see that the gravitational field (\ref{a-ch}) admits a set of solutions with
fixed $h_{22}$ as the given function of $\phi $.
More wide class of solutions will be presented in the separate article \cite{cheabb12}.

\section{Cosmological perturbations as quantum fluctuations for CCM}

Inflationary paradigm is represented by the assertion that before the epoch
of primordial nucleosynthesis the Universe expansion was accelerated\footnote{
In respect to present Universe acceleration there are division for early inflation
and later inflation}. The existence of such inflationary period gives possibility
to solve the flatness, the horizon and the monopole problems of the standard
Big Bang cosmology \cite{koltur90}. Together with this advantage inflation can
explain the production
of the first density perturbations in the early Universe which are seeds to formation
of large scale structure and for CMB temperature anisotropies. The primordial density
perturbations are created from quantum fluctuations of the scalar field (inflaton),
which is the source of the early inflation. Let us mention here that the existence of
inflationary period is strongly confirmed by observations from COBE satellite and WMAP
mission (for recent review see \cite{baupei09}). 

Our purpose now is to extend the perturbation theory from scalar singlet (inflaton)
to scalar multiplett represented by CCM. To this end
let us follow by the basic idea of inflation in respect to large
scale structure formation. Namely we suggest that quantum fluctuations of
the chiral fields as well as metric fluctuation of the internal space in CCM lead
to cosmological inhomogeneities in the same manner as the inflaton singlet
fluctuation leads to primordial density perturbations, which finally lead to
Universe structure. On this way we can use the standard results for gravitational
perturbations which can be divided for scalar, vector and tensor types. Our task then
will be to calculate the perturbations from energy momentum tensor of CCM.

Background equations for the chiral cosmological model are presented at the
section \ref{CCM}.

It is well known that vector and tensor perturbations could not
lead to instability. While the scalar perturbations give the growing
inhomogeneities which effect to the dynamics of matter.
Therefore much attention will be payed to the scalar perturbations
which can lead to the structure formation.

\subsection{General equations for chiral fields perturbations}

Let us consider the perturbations of the gravitational field with the metric
$g_{\alpha\beta}$ in a general form
\beq\label{gf-tr}
\tilde g_{\alpha\beta}=g_{\alpha\beta}+\delta g_{\alpha\beta}({\bf x},t),
~~\tilde g^{\alpha\beta}=g^{\alpha\beta}-\delta g^{\alpha\beta}({\bf x},t).
\eeq
Here $g_{\alpha\beta}(x)$ is the background metric,  $\tilde g_{\alpha\beta}$ is a
perturbed metric, $\delta g_{\alpha\beta}$ is a metric perturbation.
Note that
$\delta g_{\alpha\beta} $ can be presented as scalar, vector or tensor perturbations
of the gravitational field \cite{Lifshitz46}.

To find the equations for the perturbations of the first order we can use the
following approximation
\bear\label{cf-tr}
&\tilde \varphi^A=\varphi^A +\delta\varphi^A({\bf x},t),
\\
\label{ts-pert}
&\tilde h_{AB}(\tilde \varphi^C) \equiv h_{AB}(\tilde \varphi^C)=
h_{AB}(\varphi^C)+ h_{AB,C} \delta\varphi^C.
\ear
The last expression (\ref{ts-pert}) means that the chiral metric perturbations
are due to chiral fields perturbations only, not from perturbations internal metric form.
From (\ref{ts-pert}) one can calculate the derivative with respect to
spacetime coordinates with the formula
\beq
\partial_\alpha h_{AB}(\tilde{\varphi}^C)=h_{AB,C}\varphi^C_{,\alpha} +
h_{AB,CD}\varphi^D_{,\alpha}\delta\varphi^C
\eeq

To derive perturbed equations for chiral fields dynamic let us transform the fields
equations (\ref{4}) to the following view

\beq
\Half g_{,\alpha}g^{\alpha\beta}h_{AB}\varphi^B_{,\beta}+
g\partial_\alpha \left(g^{\alpha\beta}h_{AB}\varphi^B_{,\beta}\right)-
\Half gh_{BC,A}\varphi^B_{,\alpha}\varphi^C_{,\beta}g^{\alpha\beta}+
g\partial_AV(\varphi)=0.
\eeq
Here $g=det(g_{\alpha\beta})$.

The Einstein equation (\ref{ein-st}) for small perturbations linearized
about background metric $g_{\mu\nu}$ takes the form
\beq\label{ee-st}
\delta G^\mu_\nu = \kappa \delta T_\nu^\mu .
\eeq

Background components of the energy-momentum tensor
are presented in the formula (\ref{ch-em}).
Variation  $\delta T^\mu_\nu = \tilde T^\mu_\nu - T^\mu_\nu $
can be reduced to the form
\bear\nonumber
\delta T^\mu_\nu = \delta\varphi^A_{,\nu} \varphi_A^\mu +
\delta\varphi^B_{,\alpha}\lbrace \varphi_{B,\nu} g^{\alpha\mu}
-\delta^\mu_\nu\varphi_B^{,\alpha}\rbrace +
\delta\varphi^C \{h_{AB,C}\varphi^A_{,\nu}\varphi^B_{,\alpha} g^{\mu\alpha}-
\delta^\mu_\nu \half h_{AB,C}\varphi^A_{,\alpha}\varphi^B_{,\beta} g^{\beta\alpha}
-\delta^\mu_\nu V_{,C}\}
\label{pem}\\
-\delta g^{\mu\alpha}h_{AB}\varphi^A_{,\nu}\varphi^B_{,\alpha} +
\half\delta g^{\alpha\beta}\delta^\mu_\nu h_{AB}\varphi^A_{,\alpha}\varphi^B_{,\beta}.
\nonumber
\ear
Using relations (\ref{gf-tr})-(\ref{ts-pert}) one can decompose the chiral
fields equations (\ref{4}) into background equations and perturbed ones.
In accordance with our notations (\ref{gf-tr}) and (\ref{cf-tr}) the
background equations take the same form as the equations (\ref{4}).
The equations for the chiral fields perturbations can be transformed
to the form
$$
\delta\varphi^B_{,\beta\alpha}gg^{\alpha\beta}h_{AB}+
\delta\varphi^B_{,\beta}\lbrace\half g_{,\alpha}g^{\alpha\beta}h_{AB}
+gg^{\alpha\beta}_{,\alpha}h_{AB}+
gh_{AC,B}g^{\alpha\beta}\varphi^C_{,\alpha}+
gg^{\alpha\beta} h_{AB,C}\varphi^C_{,\alpha}-
gh_{BC,A}\varphi^C_{,\alpha}g^{\alpha\beta}\rbrace+
$$
$$
\delta\varphi^C\lbrace \left(\half g^{,\beta}
+gg^{\alpha\beta}_{,\alpha}\right)h_{AB,C}\varphi^B_{,\beta}+
gg^{\alpha\beta}h_{AB,DC}\varphi^D_{,\alpha}\varphi^B_{,\beta}+
gg^{\alpha\beta}h_{AB,C}\varphi^B_{,\beta\alpha}-
\half gg^{\alpha\beta}h_{DB,AC}\varphi^B_{,\alpha}\varphi^D_{,\beta}-
gV_{,AC}\rbrace-
$$
$$
\delta g^{\alpha\beta}\lbrace gh_{AB,C}\varphi^C_{,\alpha}\varphi^B_{,\beta}+
gh_{AB}\varphi^B_{,\alpha\beta}-\half gh_{BC,A}\varphi^B_{,\alpha}\varphi^C_{,\beta}+
\half g_{,\alpha}h_{AB}\varphi^B_{,\beta}\rbrace
-\delta g_{,\alpha}^{\alpha\beta}gh_{AB}\varphi^B_{,\beta}+
\half \xi_{,\alpha}gg^{\alpha\beta}h_{AB}\varphi  ^B_{,\beta}=0.
$$
Here $\xi=g^{\alpha\beta}\delta g_{\alpha\beta}$.

Presented in this section formulas solve the problem of construction the
theory for scalar, vector and tensor type perturbations in the chiral
cosmological model. It should be mentioned here that the solutions for
the perturbations of the gravitational field (left hand side of the equation
(\ref{ee-st})) are independent from the source and have been obtained
by Lifshitz \cite{Lifshitz46}. %

Let us turn our attention to the scalar perturbations of the gravitational
field and the perturbations of the chiral fields.

\section{Perturbations for CCM in FRW Universe}

To present the equations for cosmological perturbations we will use the conformal
time $ \eta $ instead of cosmic time $t$ as it is generally excepted
(see, for example, review by Mukhanov et al \cite{mufebr92}).
First of all let us rewrite the background Einstein and chiral field equations
in the conformal time defined by the relation $d\eta = \frac{dt}{a(t)}$.

\subsection{Background equations}

Let us consider now the chiral cosmological model (\ref{1}) in the framework
of homogeneous and isotropic FRW Universe with the metric (\ref{frw}).
The background Einstein equations (\ref{ein-st}) in terms of the conformal time
take the following form
\bear
&
3\lbrace {\cal H}^2+\epsilon\rbrace =\kappa\lbrace\Half h_{AB}\varphi^A_{,\eta}
\varphi^B_{,\eta}+a^2V \rbrace,\label{bee-1}
\\
&
a^{-2} h_{AB}\varphi^A_{,\eta}\varphi^B_{,i}=0,\label{bee-2}
\\
&
\left( 2{\cal H}^{\prime}+{\cal H}^2+\epsilon\right)=\kappa\lbrace -\Half
 h_{AB}\varphi^A_{,\eta}\varphi^B_{,\eta}+a^2V \rbrace .\label{bee-3}
\ear
The chiral fields equations (\ref{4}) become
\beq
2{\cal H}\varphi^{\prime}_A+\left(h_{AB}\varphi^{\prime B}\right)^\prime -
\Half h_{BC,A}\varphi^{\prime B}\varphi^{\prime C}+a^2V_{,A}=0.\label{bcfe}
\eeq

\subsection{Perturbed Einstein equations}

To deals with structure formation we count the scalar type perturbations and
choose the longitudinal gauge with the line element
\beq\label{lg}
ds^2=a^2(\eta)\left\{ (1+2\Phi)d\eta^2- (1-2\Phi)\gamma_{ij}dx^idx^j\right\}.
\eeq

Using the perturbations of the energy-momentum tensor $\delta T_\nu^\mu $
(\ref{pem})
and the perturbations of the gravitational field in the longitudinal gauge
(\ref{lg}) (see the equations (4.15) in review \cite{mufebr92})
one can derive the perturbed Einstein equations
\bear
\Delta \Phi-3{\cal H}\Phi^\prime -\Phi(2{\cal H}^2+{\cal H}^\prime)+
4\epsilon\Phi =
\frac{\kappa}{2}\left\{ \langle\delta\varphi_{,\eta}\varphi_{,\eta}\rangle
+\delta\varphi^C\left[ \Half h_{AB,C}\varphi^A_{,\eta}\varphi^B_{,\eta}+
a^2V_{,C}\right]\right\},\label{pee-1}\\
\left({\cal H}+\Phi^\prime \right)_{,i}=
\frac{\kappa}{2}h_{AB}\delta\varphi^A_{,i}\varphi^B_{,\eta},\label{pee-2} \\
\Phi''+ 3{\cal H}\Phi^\prime +\Phi(2{\cal H}^2+{\cal H}^\prime)=
\frac{\kappa}{2}\left\{ \langle\delta\varphi_{,\eta}\varphi_{,\eta}\rangle
+\delta\varphi^C\left[ \Half h_{AB,C}\varphi^A_{,\eta}\varphi^B_{,\eta}-
a^2V_{,C}\right]\right\},\label{pee-3}
\ear
where
$
\langle\delta\varphi_{,\eta}\varphi_{,\eta}\rangle =
h_{AB}\delta\varphi^A_{,\eta}\varphi^B_{,\eta}.
$

Actually, one can use only two perturbed Einstein
equations, because the linear combination of the two
equations (\ref{pee-2}), (\ref{pee-3}) and the
consequence of the background equations (\ref{bee-1})-(\ref{bee-3}) as well
as the fields equations (\ref{bcfe}) lead to the third equation (\ref{pee-1}).

\subsection{Perturbed chiral fields equations}

To understand the influence of the internal (target) space on the
cosmological perturbations let us insert the metric component (\ref{lg})
into the chiral fields equations (\ref{4}).
In general case the result is too long
to be presented in this review, therefore let us restrict ourself by the
case of a spatially flat Universe assuming $\epsilon=0$ in (\ref{frw}).
Finally, the chiral fields equations can be reduced to the following view
\bear 
\nonumber
(\delta \varphi_A)''-\gamma^{ij}(\delta\varphi_A)_{,ij}+2(\delta\varphi^B)'
\left\{h_{AB}{\cal H}+\Gamma_{BC,A}(\varphi^C)' \right\}+
\delta\varphi^C\left\{ \left[2{\cal H}(\varphi^B)'+(\varphi^B)''\right]h_{AB,C}+
(h_{AB,DC}-\right.\\
\nonumber
\left.\half h_{DB,AC})(\varphi^B)'(\varphi^D)'+ a^2V_{,AC}\right\}-
2\Phi \left\{\left[h_{AB,C}-\half h_{BC,A}\right]
(\varphi^B)'(\varphi^C)'+h_{AB}\left[(\varphi^B)''+
\half {\cal H}(\varphi^B)'\right]\right\} \\ %
\label{pcfe-1}
-4(\varphi^B)'h_{AB}\left(\Phi'-\Phi {\cal H}\right) =0.
\ear

It needs to emphasize now that for multiple-field models 
the chiral metric is a constant matrix. Therefore the terms
containing chiral metric derivatives will vanish in equations (\ref{pcfe-1}).
Thus the fact that the internal chiral space gives an additional income
to dynamics of cosmological perturbations is proved.
The following task will be to decompose the income of auxiliary chiral
fields (dark sector fields) and the target space metric from the leading field
(an inflaton in the most cases). Another words, it would be desirable
to safe the contribution in the cosmological perturbations from inflaton
and count separately additional contribution from dark sector chiral
fields and from the target space metric.

\section{Decomposition of the cosmological perturbations for
inflaton and dark sector ones}

Let us suppose that in the range with inflaton we have during inflationary
stage strong or weak fields of dark sector \cite{chepan10}. 
We could not make differences between types
of DS fields before equation of state analysis will be done. Also it is worth
to mention about the following understanding of the very early Universe.
If we suggest that one (say, inflaton) field exists, then it is easy to
imagine that this field has the sense of an effective field which presents
few others \cite{ch97b}. Namely
the relation between multiplett
of scalar fields with geometrical (kinetic) interaction (that is,
a chiral cosmological model, as we stressed above) can be
described by an effective singlet with the relations \cite{ch97b} 
\beq\label{nsm-ssf}
h_{AB}\varphi_{,\mu}^A\varphi_{,\nu}^B = \phi_{,\mu}\phi_{,\nu},~~
W(\varphi (x^\mu))=V(\phi (x^\mu)),~~\varphi=(\varphi^1, \ldots ,\varphi^N).
\eeq
Hereafter $W(\varphi (x^\mu))$ will be presented as the potential of
(self)interaction
for CCM when it needs to make difference from the potential of selfinteraction
$V(\phi (x^\mu))$ for the scalar singlet $\phi$.
Thus from GR point of view for CCM and selfgravitating scalar singlet
we effectively have the same type of the source
of the gravitational field, but -- more complicated dynamics in
the case of a chiral cosmological model. Moreover from (\ref{nsm-ssf}) one can
find the restriction on the chiral metric
\beq
h_{AB}=\frac{W_{,A}W_{,B}}{V'^2},~~ V'^2 \neq 0
\eeq

Our next wish is to safe the equations for cosmological perturbations for the
inflationary field as they are very successful in understanding of current observations
and very good suit them.
To this end let us decompose the scalar product in the target space by the following
way.
\beq
h_{AB}\varphi^A_{,\mu}\varphi^B_{,\nu}= h_{11}\varphi^1_{,\mu}\varphi^1_{,\nu}
+h_{1b}\varphi^1_{,\mu}\varphi^b_{,\nu}+h_{b1}\varphi^b_{,\mu}\varphi^1_{,\nu}
+h_{ab}\varphi^a_{,\mu}\varphi^b_{,\nu},\label{tsd}
\eeq
where $a,b,...$ takes the values $2,3,...N$.
Without the loss of generality we can choose gaussian coordinates with $h_{11}=\epsilon$
where $\epsilon =1$ corresponds to canonical scalar field, while $\epsilon=-1$
corresponds to phantom scalar field. (About possible phantom inflation have been
stressed in the work by \cite{canood06}. In according with gaussian coordinates we
should also set $h_{1a}=0$.

Let denote the leading field $\varphi^1$ as an inflationary field $\phi$,
that is $\varphi^1=\phi$.

To single out the inflationary perturbations from dark sector ones
let us decompose the perturbation of the
gravitation field $\Phi$ into two parts:
\beq\label{pgfd-lg}
\Phi(x,t)=\Phi_\phi +\Phi_\sigma = \Phi+\Sigma,
\eeq
where $\Phi_\phi$ is responsible for the perturbations, coming from the
inflationary field. Using the decompositions (\ref{tsd}) and (\ref{pgfd-lg})
one can separate the perturbations $\Phi$ and $\Sigma$
in the perturbed Einstein equations (\ref{pee-1})-(\ref{pee-3}) and in the
perturbed fields equation (\ref{pcfe-1}).

Einstein equations after the separation of the perturbations take the form
\bear\label{Phi-1}
\Phi''+\Delta\Phi=\kappa\phi'\delta\phi',
\\
\label{Phi-2}
\Sigma''+\Delta\Sigma=\kappa\left(h_{ab}\delta\varphi^{\prime a}\varphi^{\prime b}+
\Half \delta\varphi^C h_{ab,C} \varphi^{\prime a}\varphi^{\prime b}\right),
\\
\label{Phi-2b}
{\cal H}\Phi+\Phi'=\frac{\kappa}{2}\phi'\delta\phi,
\\
{\cal H}\Sigma+\Sigma'=\frac{\kappa}{2} h_{ab}\delta\varphi^a\varphi^b.
\ear
Hereafter we use the simplification in notations
$$
\varphi^{\prime a} := (\varphi^{a})',~~\varphi''^{b} := (\varphi^{b})''
$$
to make formulas more readable.

The perturbed chiral fields equations (\ref{pcfe-1}) can be also decomposed
and display as
\bear\label{Phi-3}
\delta\phi''-\delta^{ij}\delta\phi_{,ij}+\delta\phi'2{\cal H}+
\delta\phi a^2W_{,\phi\phi}-4\phi'\Phi'+2\Phi a^2W_{,\phi}=0,
\\
-4\phi'\Sigma'+2\Sigma a^2W_{,\phi}+\delta\varphi^b a^2W_{,b\phi}-
\delta\varphi^{\prime b}h_{ba,\phi}\varphi^{\prime a} -
\Half \delta\phi h_{ba,11}\varphi^{\prime b} \varphi^{\prime a} -
\Half \delta\varphi^d h_{ba,\phi d}\varphi^{\prime b} \varphi^{\prime a}=0,
\\
\delta\phi a^2W_{,a1}+\delta\varphi^ca^2W_{,ac}-
4h_{ab}\varphi^{\prime b}\left(\Phi'+\Sigma'\right)+
2a^2\left(\Phi+\Sigma\right)W_{,a}+ \delta\phi h_{ab,1}\varphi^{\prime b}+
\delta\varphi^{\prime b}2\Gamma_{bc,a}\varphi^{\prime c}+
\\
\delta\phi \left[ \left(2{\cal H}\varphi^{\prime b}+
\varphi''^b \right)h_{ab,1}+ \left(h_{ab,d1}-
\Half h_{db,a1} \right)\varphi'^b\varphi'^d\right]+
\\
\delta\varphi^c \left[ \left(2{\cal H}\varphi^{\prime b}+
\varphi''^b \right)h_{ab,c}+ \left(h_{ab,dc}-
\Half h_{db,ac} \right)\varphi'^b\varphi'^d +\right.
\left. h_{ab,1c}\phi'\varphi'^b\right]+ h_{ab}\left(\delta\varphi''^b-
\delta^{ij}\delta\varphi^b_{,ij} \right)+2{\cal H}\delta\varphi'^bh_{ab}=0.
\ear

The decomposition made with the aim to keep the equations for the gravitational
perturbation $\Phi$ (\ref{Phi-1}),
(\ref{Phi-2}) and (\ref{Phi-3}) in the same form as for
the scalar field (inflaton) perturbations in the longitudinal gauge (compare with the
formulas (6.40-42),(6.46) in \cite{mufebr92}).

\section{Decomposition in the two-component CCM} 

Let us consider the application of decomposition of perturbations for the two
component chiral cosmological model.
Namely, let $\varphi^1=\phi$ is the inflaton,
$\varphi^2=\chi$ is the auxiliary chiral field belonging to dark sector.
Let us mention once again that physical sense of the dark sector field we
can understand over equation of state analysis.
We set  $h_{11}=1,~h_{12}=0$ in
accordance with (\ref{tsd}). Then the background equations read
\bear
&
3\lbrace {\cal H}^2+\epsilon\rbrace =\kappa\lbrace\Half \phi'^2+\Half h_{22}\chi'^2
+a^2V \rbrace,\label{bee-1-2}
\\
&
a^{-2} \left[\phi'\phi_{,i}+h_{22}\chi'\chi_{,i}\right]=0,\label{bee-2-2}
\\
&
\left( 2{\cal H}^{\prime}+{\cal H}^2+\epsilon\right)=\kappa\lbrace -\Half
\phi'^2-\Half h_{22}\chi'^2 +a^2V \rbrace .\label{bee-3-2}
\ear
The chiral fields equations (\ref{bcfe}) are
\bear
\phi''+2{\cal H}\phi'-\Half h_{22,\phi}\chi'^2+a^2V_{,\phi}=0 \\
(h_{22}\chi')'+2{\cal H}h_{22}\chi' -
\Half h_{22,\chi}\chi'^2+a^2V_{,\chi}=0.\label{bcfe-2}
\ear

Let us suggest for the sake of simplicity that
$h_{22}$ is the function depending on
the leading field only: $h_{22}=h_{22}(\phi)$.
Let the potential $V$ will obey the same property: $V=V(\phi)$.

Under the assumptions above one can obtain the consequences from Einstein
equations (\ref{bee-1-2})-(\ref{bee-3-2}):
\beq\label{cbee-1}
V=\frac{\left( {\cal H}'+2{\cal H}^2\right)}{\kappa a^2},~
(\phi')^2+h_{22}(\chi')^2=\frac{2\left({\cal H}^2- {\cal H}'\right)}{\kappa}.
\eeq

The equations (\ref{cbee-1}) give us possibility to apply the method of
chiral space metric deformation \cite{Chervon95a}, \cite{Chervon95b} 
for construction solutions for cosmological perturbations.
The general scheme can be described as follow. One should specify the regime
of scalar factor evolution $a=a(\eta)$, then from the first equation of
(\ref{cbee-1})
one can find $V=V(\eta)$ as well as the linear combination of the left hand side
of the second equation of (\ref{cbee-1}). Using 
(\ref{bcfe-2}) one can find the relation between inflationary field $\phi$ and
a chiral metric component $h_{22}$:
\beq
(\phi')^2={\cal F}(\eta) -\frac{C_1}{h_{22}a^4},
~{\cal F}(\eta)=\frac{2}{\kappa}\left({\cal H}^2- {\cal H}'\right).
\eeq
Here $C_1=\const$ is the integrating constant.

As it was pointed out earlier the inflationary field perturbations exactly
extracted from the perturbed CCM equations. In the case under consideration one can
derive the equation for auxiliary dark sector field perturbations. The equation is simplified
if make the additional suggestion about vanishing of the second term in r.h.s. of
(\ref{Phi-2}). To this end we can restricted consideration for the time when
chiral metric component $ h_{22}$ takes it maximum value. Then the equation for the dark
sector field (\ref{Phi-2}) simplified to the view:
\beq\label{nip}
\Sigma''-\Delta\Sigma +6{\cal H}\Sigma'+2\Sigma\left({\cal H}'+
2{\cal H}^2\right)=0.
\eeq
That is the equation above entirely determines the gravitational perturbation
$ \Sigma $ from the dark sector field $\chi$.

Let us transform the equation (\ref{nip}) to more suitable form. For this
purpose we introduce new variable $N$ by the following manner
\beq\label{v-N}
\Sigma = \frac{N}{a^3(\eta)},~~N=N(\eta , x^i).
\eeq

The equation (\ref{nip}) then reduced to the following one
\beq\label{nip-N}
N''-\Delta N-N (5{\cal H}^2+{\cal H}')=0.
\eeq
Let us try to find the solutions in the standard form
$N=Y(\eta) \exp\{i\bar k\bar x\}$. The equation (\ref{nip-N}) takes the form
\beq\label{nip-Y}
Y''+Y(\eta)k^2+Y(\eta)(5{\cal H}^2+{\cal H}')=0.
\eeq

This equation can be easily solved in the shortwave approximation, when
$k^2 \gg (5{\cal H}^2+{\cal H}')$. The solution is
\beq
Y=\pm \frac{C}{k}\sin k(\eta+\eta_0)
\eeq
Finally, using the relation (\ref{v-N}), one can obtain
\beq
\Sigma=\pm \frac{C}{a^3k}\sin k(\eta+\eta_0)\exp\{i\bar k\bar x\}
\eeq
This result can be also presented in the form
\beq\label{nonip-1}
\Sigma=\pm \frac{C}{a^3k}\frac{1}{2i}\left(e^{\{i k(\eta+\eta_0)\}}-
e^{\{-i k(\eta+\eta_0)\}} \right)\exp\{-i\bar k\bar x\}
\eeq
Let us compare obtained solution for dark sector field perturbations with
inflaton ones, described by formula
\beq\label{inp-gf}
\Phi=\pm \frac{C_1}{a k}\frac{1}{2i}\left(e^{\{i k(\eta+\eta_0)\}}-
e^{\{-i k(\eta+\eta_0)\}} \right)\exp\{-i\bar k\bar x\}.
\eeq
We can conclude that dark sector field perturbations decay faster (if $a >1$)
then the inflaton ones during expansion of the Universe.

\section{Decomposition of perturbations for power law Universe \\ expansion}

The background solution for the power law inflation  $a = a_{0} t^{m}, m>1$
can be found by the fine turning method and has the form
\cite{chezak97}, \cite{chekos03}.

\bear
&&\chi = 2C_0 \int {\frac{dt}{h_{22} \left( \phi (t) \right)t^{3m}}},\\
&&\phi = \int { \sqrt {\frac{2m}{ \kappa t^2} - \frac{C_0^2}
{h_{22} \left( \phi (t) \right)t^{6m}}} dt},\\
&&W\left( \phi (t) \right) = \frac{ m(3m - 1) }{ \kappa t^2 }.
\ear

General case of study small perturbations is too difficult for investigation.
Therefore we choose the simplification with the time dependence
for the chiral metric in the form
$h_{22}(t) = \frac{2}{A}t^{2 - 6m}$, where $A$ is a positive constant: $A>0$.
From this suggestion one can find

\bear
&h_{22}(\phi)=\frac{2}{A}\exp\{(2-6m)\sqrt{\frac{\kappa}{2(m-d)} }\phi \},\\
&W(\phi)=\frac{m(3m-1)}{\ae}\exp\lbrace -2\sqrt{\frac{\kappa}{2(m-d)}}\phi\rbrace,
\ear
where $d=\kappa A C_0^2$.

New approach for solving the complete system of Einstein and fields perturbed equations
was proposed by \cite{Koshelev01}. The system of above mentioned equations was reduced
to the ordinary differential equation of the forth order

\bear\nonumber
&& \Phi ^{(4)} + 2\frac{2m + 3}{t}\Phi ^{(3)}
+ (\frac{\Pi_1(m,d)}{(m-d)t^2} + 2\frac{k^2}{a_0^2}t^{-2m})\ddot{\Phi}\\
\nonumber
&&+(3\frac{m(2+m)(6md+m-3d)}{(m-d)t^3}+6\frac{k^2}{a_0^2}t^{-2m-1})\dot{\Phi}+\\
\label{eq11}
&&(\frac{(\Pi_2(m,d)k^2}{(m-d)a_0^2}t^{-2m-2}+\frac{k^4}{a_0^4}t^{-4m})\Phi=0,
\ear
where $\Pi_1 (m,d) = 3m^3 + 15m^2 d - 20md + 14m^2 + 6m - 6d,
~~\Pi_2(m,d) =- 2m^3 - 16m^2d + 18md + 2m - 6d,~~C_0\ne 0$.
The equation (\ref{eq11}) obeys by the property that $\Phi $ and calculated with it
$\delta \chi $ are the solutions of the complete system of perturbed equations.
If we know the perturbation of the gravitational field $\Phi $,
the value of $\delta \varphi $ and $\delta \chi $ can be easily found.

In the longwave approximation, when the size of nonhomogeneities is much more
then the size of a horizon, one can suggest
$\frac{k}{a_0 t^{m - 1}} \ll 1$. Under this suggestion we obtain the Euler
equation and the fist of its solutions is evident: $\Phi _k = const $.
The second one has the form $\Phi _k = const \cdot t^{- m - 1}$.
The third and forth independent solutions are depend on the value of the
parameters {\it m} and {\it d}. When $0 < d < m$ the later solutions are fast
decreasing ones. For example, when $m = 3$ and $d =1$ the complete solution
takes the form

\begin{equation}
\label{eq12}
{\begin{array}{*{20}c}
{\Phi _k = A_k + B_k t^{ - 4} }
  \hfill 
{+ C_k t^{ - 4}\cos(2 \sqrt {14} \ln t )
 + D_k t^{ - 4}\sin(2 \sqrt {14} \ln t ) }
  \hfill \\
\end{array} }
\end{equation}

If we restrict ourself only by nondecreasing mode of the metric perturbation
then the chiral fields dependence on time takes the form
$\delta \varphi = const$, $ \delta \chi \propto t^{3m-1}$,
besides the relation
$ \frac{\delta \varphi}{\dot{\varphi}}= \frac{\delta \chi}{\dot{\chi}}$
is valid.

The growing modes of metric perturbations may appear in the case of negative
values of {\it d }, i.e. during the period when the potential $W$ has negative
values.

In the shortwave approximation the solution of the equation of the forth
order 
is

\bear\label{shw-1}
&&\Phi _k^{(1)} = \frac{\exp( {\pm \frac{ik}{a_0} \frac{t^{1 - m}}{1-m} )}}{t}\\
\label{shw-2}
&&\Phi _k^{(2)} = \frac{\exp( \pm \frac{ik}{a_0} \frac{t^{1 - m}}{1-m} )}{t^{3m}}
\ear

Corresponding to the solution above the chiral fields perturbations take
the form

\bear
&&\delta \chi = \pm \frac{d}{m \kappa C_0} \frac{ik}{a_0}t^{2m-1}
\exp( \pm \frac{ik}{a_0} \frac{t^{1 - m}}{1-m} ),\\
&&\delta \varphi = \pm \frac{1}{m} \sqrt{ \frac{2m-2d}{\kappa}} \frac{ik}{a_0}
t^{-m}\exp( \pm \frac{ik}{a_0} \frac{t^{1 - m}}{1-m} )
\ear

One can see, that the metric perturbations $ \Phi$ of the form (\ref{shw-1})
are coincident with the case of the power law inflation with one inflaton
field. Besides the perturbations of $ \delta \varphi$ are differ from
inflaton ones only by a constant factor and the relation
$$
\frac{\delta \varphi}{\dot{\varphi}}= \frac{\delta \chi}{\dot{\chi}}
$$
is valid.

Thus the components of the model under consideration are strongly coupled
even in the short wave approximation.

For the fast decreasing modes (\ref{shw-2}) the perturbations of
$\delta \varphi$
can be considered in the framework of used approximation as zero.
While for $\delta \chi$ one can obtain the expression
$$
\delta \chi = \frac{1}{\kappa C_0} \frac{ik}{a_0}t^{-m}
exp( \pm \frac{ik}{a_0} \frac{t^{1 - m}}{1-m} ).
$$
It should be mentioned also about strong correspondence of two approach
for cosmological perturbations. Solutions (\ref{shw-1},\ref{shw-2}) are
obtained by solving the equation of the forth order, while solutions
(\ref{nonip-1},\ref{inp-gf}) are obtained directly from decomposed equations.

\section{Dark sector fields on the inflationary background}

In previous section we analyzed decomposition of the chiral fields for inflaton and
dark sector fields in the framework of self-gravitating NSM with the potential
of (self)interaction. It may happened that the dark sector fields were formed earlier
then inflation starts. In this case we may think over new scenario in which dark sector fields
are very weak and do not affect for gravitational field supported by inflaton.
Analogical situation was studied by Jaffe \cite{Jaffe94} who modified and extended
formalism
that have been developed by Veeraraghavan and Stebbins \cite{veeste89}.
The approach was based
on the perturbations produced by a "stiff source" such as these cosmic field configuration
formed by Kibble mechanism. A "stiff source" evolves in the background of FRW Universe;
the back reaction of gravitational perturbations onto the source is regarded to be negligible.
In our approach \cite{chepan10}, \cite{panche11} we prefer to use the term "weak source"
instead of
"stiff source". Note that during inflation we assume the weakness of the chiral fields
$\varphi^C$ in respect to inflaton $\phi $; also we regard the magnitude
of dark sector
fields are on the same level as for perturbed inflation field $\delta\phi .$

\subsection{Basic equations of the model}
To describe weak dark sector fields on the inflationary background let us construct the
action integral

\beq\label{gen-syst}
S_w=\int\sqrt{-g}d^4x\left[\frac{R-2\Lambda}{2\kappa}+\frac12
\phi_{,\mu}\phi_{,\nu}g^{\mu\nu}-V(\phi)
  +\frac12 h_{AB}(\varphi)\varphi_{,\mu}^A
\varphi_{,\nu}^B g^{\mu\nu}-W(\varphi)\right].
 \eeq
Here we denoted the leading field -- the inflaton as $\phi $; the
chiral fields $ \varphi^C $ describe dark sector fields.

For the sake of simplicity let us start with two dark sector
fields $\varphi^1= \psi, ~ \varphi^2 =\chi $. This case will be
represented by the two component CCM and we choose the metric in
gaussian coordinates
$$
d\sigma^{2}= d\psi^{2}+ h_{22}(\psi,\chi) d\chi^{2}.
$$

To describe the physical situation we should introduce some
restrictions on the potential and kinetic energies of the fields
under consideration. To this end let us represent the total
potential as the sum of two parts:
$$W_{tot}(\phi, \psi, \chi) = V(\phi) +
W(\psi, \chi).$$
Besides we count that $ W(\psi, \chi)\ll V(\phi)$, therefore
$W_{tot}\approx V(\phi)$.

Similar relations we set for inflaton (leading field) kinetic energy
$K_{L}(\phi)= \frac12 \phi_{,\alpha}
\phi^{,\alpha}$ and dark sector fields kinetic energy 
$ K_{12}(\psi, \chi)=\frac12
\psi_{,\alpha}\psi^{,\alpha}+\frac12
h_{22}(\psi,\chi)\chi_{,\alpha}\chi^{,\alpha}:$

$$K_{tot}(\phi,
\psi, \chi)=K_{L}(\phi)+K_{12}(\psi, \chi).$$
Supposed that $ K_{12} \ll
K_{tot}, $ we obtain $K_{tot} \approx K_{L}$.

After the restrictions above the kinetic energy term $K_{12}$ and the
potential $W(\psi,\chi)$ of CCM should be deleted from Einstein equations
during the inflationary period.
In terms of the conformal time for FRW Universe with the metric
\begin{equation}\label{frw}%
ds^{2}=a^2(\eta)(-d\eta^{2}+\gamma_{ij}dx^idx^j), %
\end{equation}%
(where $\gamma_{ij}$ being the metric of three dimensional space-like section)
the equations (\ref{bee-1}), (\ref{bee-3}) take the following form
\bear
\label{E_3-1}
3\mathcal{H}^2=\kappa \left(\mathcal{K}_L+a^2 V(\phi)\right), \\
\label{E_3-2}
\mathcal{H}'-\mathcal{H}^2=-\kappa \mathcal{K}_L,  %
\ear
where $ \mathcal{H} $ = $ \displaystyle\frac{a'}{a} $ and $\mathcal{K}_L = \Half {\phi'}^2. $

Dynamic equation for inflaton reduced to \cite{mufebr92}:
\begin{equation}\label{fe0} %
\phi''+2 \mathcal{H} \phi'+a^2 V_{, \phi} = 0.%
\end{equation} %

Thus, obtained equations (\ref{E_3-1})-(\ref{fe0}) represent the
inflationary epoch driving by inflaton $ \phi $ with the potential
of selfinteraction $V(\phi)$. Note that the equations
(\ref{E_3-1}), (\ref{E_3-2}) are corresponded to the Einstein
equations (\ref{E1}), (\ref{E2}) with $\epsilon =0, ~h_{11}=1,
~h_{12}=0.$

Dark sector fields equations under restrictions $ W(\psi, \chi)\ll
V(\phi)$ and $ K_{12} \ll K_{tot}, $ will account only
gravitational interaction by means of the Hubble parameter
$\mathcal{H}$ and read
\bear\label{dsf-1}
\psi''+2\mathcal{H}\psi'-\Half \frac{\partial h_{22}}{d\psi}\chi'^2+
a^2 W_{,\psi}=0, \\
\label{dsf-2}
h_{22}(\chi''+2\mathcal{H}\chi')+\frac{\partial h_{22}}{d\psi}\psi'\chi'
+\Half \frac{\partial h_{22}}{d\chi}\chi'^2 + a^2 W_{,\chi}=0. %
\ear

The simplification for the case when $h_{22}=h_{22}(\psi)$ and $W=W(\psi)$ usually
give possibility to find the examples of exact solutions for the model with
$h_{22}=\sin^2 \psi , h_{22}=\psi^2 $, etc. as it was shown at the section \ref{gqm}.

\subsection{The dark sector fields influence on the perturbations}

According with our suggestions the Einstein equations for the perturbations
can be written as

\begin{equation}\label{dee}%
\delta G_{\mu}^{\nu} = \kappa(\delta T_{\mu}^{\nu} + \Theta_{\mu}^{\nu}).%
\end{equation}%
Here
\begin{equation}\label{Tmunu} T_{\mu\nu}=\phi_{,\mu} \phi_{,\nu} -
g_{\mu\nu}\left(\frac12 \phi_{,\alpha} \phi^{,\alpha} - V(\phi)
\right),
\end{equation}
\beq\label{Theta} 
\Theta_{\mu\nu}=\psi_{,\mu}\psi_{,\nu} +
h_{22}\chi_{,\mu}\chi_{,\nu} -
 g_{\mu\nu}\left( \frac12
\psi_{,\alpha}\psi^{,\alpha}+\frac12
h_{22}\chi_{,\alpha}\chi^{,\alpha}-W(\psi, \chi)\right).
\eeq

The inflationary stage of the Universe is governed by the inflaton
having the perturbation of the first order
$\delta \phi $: $ \tilde{\phi} =
\phi + \delta \phi ,$ where $ \phi $ is the solution of the background equations
(\ref{fe0}). The corresponding perturbations of the energy-momentum tensor
(\ref{Tmunu}) $ \delta T_{\mu}^{\nu} $ are the same order of magnitude as
the energy-momentum tensor for the weak source $\Theta_{\mu\nu} $ (\ref{Theta})
for the model under investigation.

Once again we choose the longitudinal gauge with the line element (\ref{lg})
\beq\nonumber  
ds^2=a^2(\eta)\left\{ (1+2\Phi)d\eta^2- (1-2\Phi)\gamma_{ij}dx^idx^j\right\}.
\eeq

Here $ \Phi = \Phi(\eta, \vec{r}) $ is the perturbation of the metric
(of the gravitational field). Now form the equation (\ref{dee}), both parts of which
are gauge invariant \cite{mufebr92}, using the perturbed Einstein equations
(\ref{pee-1})-(\ref{pee-3}) one can obtain modified equations for longitudinal gauge
%
\begin{eqnarray}\label{dee1}
\nabla^2 \Phi -3 \mathcal{H} \Phi '
-(\mathcal{H}'+2\mathcal{H}^2)\Phi =
\frac{\kappa}{2}(\phi' \delta \phi ' + a^2 V_{, \phi} \delta \phi+\Theta_0^0),\\ %
\label{dee2}%
\Phi ' + \mathcal{H} \Phi =\frac{\kappa}{2} \left(\phi' \delta \phi +\Theta^0_*\right),\\ %
\label{dee3}
\Phi '' +3 \mathcal{H} \Phi ' +(\mathcal{H}'+ 2\mathcal{H}^2)\Phi
= \frac{\kappa}{2}(\phi' \delta \phi ' - a^2 V_{, \phi} \delta \phi-\Theta_*^*),%
\end{eqnarray}%
where $ \phi $ is the background solution of the equations (\ref{E_3-1})-(\ref{fe0})
 for the inflaton; $ \delta \phi = \delta
\phi(\eta, \vec{r}) $ is the inflaton perturbation.  %
The components of the weak energy-momentum tensor $ \Theta_\mu^\nu $ are
 $\Theta_0^0 =\rho= a^{-2}K_{12} +W $, $ \Theta_*^*=\Theta_1^1=\Theta_2^2=\Theta_3^3 =-p=
-a^{-2}K_{12} +W $.

By adding the equation (\ref{dee1}) with (\ref{dee3}) and using
consequence of background equations (\ref{bee-1})-(\ref{bee-3})
$ \mathcal{H}^2 - \mathcal{H}' = \frac{\kappa}{2} \phi'^2 $ and the relation
$ \Theta_0^0+\Theta_*^* =
\rho - p = 2W, $
one can obtain the differential equation in the partial derivatives
of the second order in respect to the function $\Phi $
\begin{equation}\label{Phi}%
\Phi''-\nabla^2
\Phi+2\Phi'\left(\mathcal{H}-\frac{\phi''}{\phi'}\right)
+2\Phi\left(\mathcal{H}'
-\mathcal{H}\frac{\phi''}{\phi'}\right)+\kappa W=0.%
\end{equation}%
Here $ \phi $ and $\mathcal{H} $ can be obtained from background
equations for the inflationary stage (\ref{E_3-1})-(\ref{fe0}).
The potential $ W $ reflects the property of the dark sector
fields and may be tested for different HEP predictions. Afterward
we can define the perturbation of the gravitational field $\Phi $
by solving the equation (\ref{Phi}) and corresponded perturbation
of the inflaton $ \delta \phi $ from
(\ref{dee2}): %
\begin{equation}\label{dphi}%
\delta \phi = \frac{2}{\kappa \phi'} (\Phi ' + \mathcal{H} \Phi) .%
\end{equation}%

Note that the inflaton perturbation $ \delta \phi $ should satisfy the perturbed
dynamic equation
\begin{equation}\label{df}%
\delta \phi''+2 \mathcal{H} \delta \phi'-
\nabla^2\delta\phi+a^2V_{,\phi\phi}\delta\phi = 0.%
\end{equation}

Now we are ready to investigate the solution for inflationary Universe with exponential
scale factor, which is very important for inflationary epoch analysis.

\subsection{The solutions on exponential inflation background}

Let us start from the given scale factor as function on time. In
according with the fine tuning method \cite{chzhsh97} we
can define the shape of the potential $V$ and the dynamics of the
inflationary field $\phi $. It is clear that this method can be
applied for the model in terms of conformal time. Therefore we set
the scale factor as the exponent of the cosmic time $t$ : $
a(t)=a_se^{h_*t} (a_s, h_* $ -- constants) and make conversion to
conformal time $\eta $.
For the conformal time we obtain
$ ~a(\eta)=-\displaystyle\frac{1}{h_*\eta} , ~ \eta =
-\displaystyle\frac{e^{-h_*t}}{a_sh_*} $.
The solutions of the model equations (\ref{E_3-1})-(\ref{fe0}) are
\beq\label{exp_1}
\mathcal{H}(\eta) = -\displaystyle\frac{1}{\eta},
 ~ \phi=const, ~ V(\phi)=const.
\eeq
Note the restriction on the conformal time $\eta : \eta \in
(-1/[a_s h_*],0)$, which is corresponded to variation of the
cosmic time ~$t: ~t \in (0,\infty).$

Taking into account the solutions (\ref{exp_1}) the modified equations
for perturbations (\ref{dee1})-(\ref{dee3}) may be reduced to
\begin{equation}\label{Phi-exp}%
\Phi''-\nabla^2 \Phi+\kappa W(\eta)=0.%
\end{equation}%
Here we follow by the procedure suggested in \cite{mufebr92}
when we consider the consequences of perturbed gravitational
equations instead of dynamical fields equations.

It was discussed in the Introduction about pure data about kinetic
and potential interactions of the dark sector fields. Therefore we
will study possible simple assumptions for $h_{22}$ and $W$ to
obtain the solutions for cosmological perturbations.

Let as set the kinetic interaction $h_{22}$ by the dependence on $\psi :
~ h_{22}= \psi^2 $ and assume the dependence of the potential $W$
only on $\psi $ by the following Higgs-type manner
\cite{panche11}

\begin{equation}\label{pot-w}
W(\psi) = \frac{h_*^2\gamma^2}{4\beta^2} \psi^4+h_*^2
\psi^2+W_*.
\end{equation}
For this type potential the solutions for the background dark sector fields evolution
read: $$ \psi = \beta \eta ,~~
~\chi = \gamma \eta + const. $$ %

This solutions give us possibility to present the potential and kinetic energy
in terms of conformal time $\eta $

\begin{eqnarray}\label{W-ot-t}
W(\eta) = h_*^2\beta^2 \eta^2\left(1 +
\displaystyle\frac{\gamma^2}{4}\eta^2\right)+W_*, \\
\label{K-ot-t}
K(\eta)=\frac12 h_*^2\beta^2 \eta^2(1 + \gamma^2
\eta^2).
\end{eqnarray} %
Here $ W_*, \beta, \gamma $ -- integration constants
Using the relation (\ref{W-ot-t}) we can solve
the equation (\ref{Phi-exp}) for the longwave approximation.
That is we in usual way represent the gravitational perturbation
as the plane waives with the wave vector ${\bf k}$ and amplitude
$\Phi$ depending on $\eta : $

$$
\Phi(\eta, \textbf{x}) = \Phi(\eta) e^{ i \textbf{kx} }.
$$
Then the term $ \nabla^2 \Phi $ can be neglected in the longwave approximation
as the low waive number $k=|{\bf k}|$. After this procedure we can solve the equation
(\ref{Phi-exp}) exactly
\begin{equation}\label{Phi-1}%
\Phi_1 = C_2 \eta - \frac{\kappa (h_*\beta\gamma)^2}{120} \eta^6
-\frac{\kappa (h_*\beta)^2}{12} \eta^4 - \frac{\kappa W_*}{2}
\eta^2.
\end{equation}

Now we can compare the result (\ref{Phi-1}) with the solution for the inflaton
perturbations
\begin{eqnarray}\label{Phi-0}%
\Phi ^{(0)} = C_2 \eta, \qquad \delta \phi ^{(0)} = C_1 \eta
^3+C_3.
\end{eqnarray}%

without the dark sector fields with the energy momentum tensor
$\Theta_\mu^\nu $.
Note that this solution will be agreed with the equation
(\ref{dee2}) if we require $W_*=0, ~\beta \ll 1$ and $\beta
\rightarrow 0$ при $\eta \simeq \eta_{infl}$.

More solutions, analysis of their physical properties and graphical illustrations were
discussed in the works \cite{chepan10}, \cite{panche11}.
Note that we used
the direct integration of the equation (\ref{Phi}) instead of the reducing this equation to
new variable \cite{mufebr92} $ u = \frac{a}{\phi'}\Phi .$ The reason is the specific property
of the investigated solution where $ \phi' =0 $ in the denominator. To get round this problem
let us consider the power law inflation.

\subsection{The solutions on power law inflation background}

Let us choose the scale factor in the power law expansion form $a=a_* t^m $ for
the cosmic time $t$. Corresponded evolution for the conformal time $\eta $
may be represented
as $a= a_s \eta^\alpha $ with $ \alpha=\frac{m}{1-m}$. Suggesting the inflation expansion
we have to set $m>1$ implying
$\alpha <-1$. Using the relations between cosmic and conformal times:
$ \eta =\frac{t^{1-m}}{a_s(1-m)},~t=[a_s(1-m)\eta]^{\frac{1}{1-m}} $ one can see
that when $t \rightarrow 0, \eta \rightarrow -\infty $ and when
$t \rightarrow \infty, \eta \rightarrow 0. $ The background solutions of the equations
(\ref{E_3-1})-(\ref{fe0}) are

\beq\label{powl_1}
\mathcal{H}(\eta) = \displaystyle\frac{\alpha}{\eta},
 ~~ \phi=\sqrt{\frac{2}{\kappa}\alpha(\alpha+1)}\ln \eta +\phi_0, ~~
 V(\phi)= \frac{\alpha (2\alpha -1)}{a_s \kappa}\exp
 \left(- \frac{(\alpha +2)(\phi -\phi_0)}{\sqrt{\frac{2}{\kappa}\alpha(\alpha
 +1)}}\right).
\eeq

Taking into account the solutions (\ref{powl_1}) one can reduce the main equation for
the gravitational perturbation (\ref{Phi}) to the following form
\beq\label{Phi_3}
\Phi''-\nabla^2
\Phi+2\Phi'\left(\frac{\alpha +1}{\eta}\right)
+\kappa W=0.%
\eeq

The potential $W$ of interaction between dark sector fields $\psi $ and $\chi $ can not
 be suggested from observation data. Therefore we can consider various functional dependence
$W$ on $\psi $ follow by intuition or analogies from particle physics for example.
Let as once again set the kinetic interaction $h_{22}$ by the dependence on $\psi :
~ h_{22}= \psi^2 $ and assume that the potential $W$ is the function
only on $\psi $ by the following type
\beq\label{w-plo} W=-K_1^2
\psi^{\lambda_1}+K_2^2\psi^{\lambda_2}+W_0.
\eeq
Here $ K_1, K_2, \lambda_1$ and $\lambda_2$ are constants which
should be matched
with the model parameters. 
Taking as in previous case the linear dependence on conformal time
for the dark sector field $\psi :~\psi=\beta \eta $ one can obtain
the solution for the second field $\chi $. Thus the solutions for
the dark sector fields are: \beq \psi=\beta \eta , ~~ \chi=\chi_0
- \frac{1}{(2\alpha+1)}\frac{1}{\eta^{(2\alpha+1)}}.
\eeq

With this solution it is not difficult to obtain the following
values for the constants \beq
K_1^2=\frac{\beta^{2(\alpha+1)}}{a_s^2},~~\lambda_1=-2\alpha,~~
K_2^2=\frac{2C_1^2\beta^{6\alpha}}{a_s^2(1-2\alpha)},~~\lambda_2=1-2\alpha
.
\eeq

Neglecting by the term $ \nabla^2 \Phi $ in the long-wavelength
approximation we can solve the equation (\ref{Phi_3}) exactly. The
gravitational perturbation is
\beq \Phi =
\frac{\kappa\beta^2}{6a_s^2(1-\alpha)}\eta^{2(1-\alpha)}
-\frac{2\kappa
C_1^2\beta^{4\alpha+1}}{4a_s^2(1-2\alpha)(3-2\alpha)}\eta^{(3-2\alpha)}
-C_2\frac{\eta^{-(2\alpha+1)}}{(2\alpha+1)}.
\eeq

It is clear that if we put suitable values for power law inflation
$m$ for cosmic time and then by calculating the power $\alpha $
for conformal time we obtain the potential of dark sector field
$W$ in terms of double power law combination (\ref{w-plo}). For
example taking $m=3$ one can obtain the potential of dark sector
fields
\beq
W= -\frac{1}{a_s^2 \beta}\psi^3+ \frac{C_1^2}{2a_s^2 \beta^9}
\psi^4+W_0.
\eeq

The gravitational perturbation (\ref{Phi_3}) reads
\beq
\Phi = \frac{\kappa\beta^2}{15a_s^2}\eta^5 -\frac{\kappa
C_1^2}{48a_s^2 \beta^5}\eta^6 +\frac{C_2}{2\eta^{2}}.
\eeq

Thus we can see the influence of the expansion parameter $ \alpha
$ and dark sector field parameter $ \beta $ for the gravitational
perturbation and the potential $W$. Of course if we may obtain
some information about dark potential $W$ we immediately can
insert it in the equations (\ref{dsf-1})-(\ref{dsf-2}) and try to
solve them.

The comparison with standard solution for long-wavelength
approximation (the formula (5.65) in \cite{mufebr92})
\beq\label{Phi_st}
\Phi^{(0)} = A\eta^{1+\alpha}, ~A= const
\eeq
stress the more crucial difference then in the case with
exponential inflation. Namely there are no terms in (\ref{Phi_3})
corresponding to standard solution (\ref{Phi_st}) for any admitted
$\alpha $. Thus taking into account the dark sector fields
influence for gravitational perturbation we may obtain a very
different picture for structure formation.


\section{Conclusions}

In the present review we show the advantage of the chiral cosmological model
for description of the early and later inflation.
Namely CCM successively used for the dark sector fields description for the
explanation of accelerated Universe expansion at the present time.

To connect the early inflation with predictions for observation
we presented general formulas for
cosmological perturbations from CCM and described the method of
decomposition of cosmological perturbations for inflaton and dark
sector ones. Based on this approach we proposed new generalized
quintom model and founded out new exact solutions for it.
We discussed new point of view on decomposition of CCM perturbations in the light of
dark energy fields using the example of power law Universe expansion.

Let us mention that the chiral cosmological model, based on the NSM with the
(self)interacting
potential, contains self-interacted single-field theories and
multiple scalar fields models which may be inspired by superstrings cosmology.
That is the chiral cosmological model can be reduced to the models mentioned
above under the special restrictions on the target space metric.

We show the new possibilities of investigation of cosmological perturbations
taking into account influence of dark sector fields on inflation background.

\end{document}